# Comment on "K. Hansen, Int. J. Mass Spectrom. 399-400 (2016) 51"

Leif Holmlid
Atmospheric Science, Department of Chemistry and Molecular Biology, University of Gothenburg, SE-412 96 Göteborg, Sweden.
Tel.:+46-31-7869076. E-mail address: holmlid@chem.gu.se

This comment was rejected by Int. J. Mass Spetrom. 2016-06-13.

The re-analysis by K. Hansen [1] of the results in one table (Table 1) in our Ref. [2] concerns a small part of that study, and does not give any new results that have not been taken into account in the analysis when completing Ref. [2]. It is important to not make fast and faulty conclusions in this kind of quite complex system: a simple analysis does not mean a correct analysis. So let us first follow the suggested re-analysis, to see what changes should be made in the conclusions of Ref. [2] if the re-analysis would be valid. The replotting in Ref [1] indicates that the KER (kinetic energy release from the Coulomb explosion) is 130 eV for p, which differs from the value of 630 eV for D from excitation level $s = 2$ used in Ref. [2]. This re-analysis [1] thus also gives a large KER > 100 eV. This is the important point in proving that the bond distances in ultra-dense hydrogen H(0) have a size of a few pm. In fact, the 130 eV value per p suggested by the re-analysis agrees quite well with the KER given to the two protons in a p-p bond in ultra-dense protium with $s = 3$ of 144 eV, thus with a p-p distance of 5.0 pm. Thus, the re-analysis in Ref. [1] gives the same type of conclusion as in our Ref. [2], that ultra-dense hydrogen exists with pm-sized interatomic distances.

The state which best matches the re-analysis [1], namely spin excitation state $s = 3$ in p(0), was in fact studied in Ref. [2] with the same TOF methods as used for D(0). (The description of ultra-dense hydrogen was recently changed from H(-1) as used in Ref. [2] to H(0)). Such results are shown in Fig. 7 and Table 3 in Ref. [2]. As can be seen there, the measured TOF times are much shorter for p(0) than for D(0) and do not agree with the re-analysis. It is thus concluded that p ejection from p(0) as suggested in Ref. [1] for Table 3 in [2] is not correct since it does not agree with measurements.

The rotational excitation implied in Ref. [2] during the Coulomb explosion to understand the low translational kinetic energy may also be discussed. In Ref. [1] this effect is supposed to be caused by the use of the atom mass D (which is from the gas admitted) instead of p (assumed in [1] to be an impurity) in the analysis in Ref. [2]. However, the same type of rotational excitation analysis has been used successfully previously also for studies of p(0) as in Ref. [3] (Table 1 and Fig. 3; Table 3 and Fig. 8). In such a case, p is already there and no lighter impurity can exist to support a re-analysis. A re-analysis of the p$^+$ data in Table 3 in Ref. [3] is shown in Fig. 1, of the same form as in Ref. [1]. It is seen that a large KER exists also in this case. The best-fit linear KER value is 170 eV, and the best-fit mass is 0.60 u, thus close to the kaon mass. However, it is unlikely that kaons are formed with such a low kinetic energy. Indeed, this plot shows that the procedure used in [1] consistently gives a too low mass. The curvature of the data in Fig. 1 is similar to that in Ref. [1], and this indicates that a more complex process like rotational excitation is needed. A full interpretation was already given in

Ref. [3]. Thus the rotational excitation effect is observed also for p(0) and is not related to any error in the mass assignment of the peaks.

This rotational effect was first introduced in Ref. [4] (Table 2, Fig. 3), where even another type of source for H(0) was employed. It was explained there that the rotational energy excited the D-D pair rotation in the clusters, of course not in an atom as believed in Ref. [1]. It may not be immediately clear how the laser-target bias potential can influence the properties of the cluster fragments as suggested in [2]. This point was however already treated in Ref. [4], section 4.2, and attributed to shorter H(0) clusters at high field strengths giving less rotational energy transfer during the Coulomb explosion. Such a shortening effect is often observed in the experiments. A sketch of a possible rotational excitation mechanism is shown in Fig. 2. Note that the clusters were shown to stand vertically on a metal surface in Ref. [5]. Due to the dynamics of the energy release in Fig. 2, a rotational excitation must in fact be assumed, and the absence of such an effect would need a good explanation. Thus, the rotational excitation observed in Refs. [2-4] is indeed expected. Recent results (submitted) give excellent pure rotational spectra for D(0) in emission in the visible range, strongly confirming the cluster rotation as used in [2] and also giving the D-D distances within 0.003 pm in the D(0) clusters. In fact, rotational motion is observed both for single and double D-D pairs.

It is also interesting to discuss if it is possible to have a source of p in the experiments as the re-analysis in Ref. [1] suggests as "a contamination of the source materials". The experiments are performed with a constant hydrogen or deuterium gas flow into the chamber under fast pumping (typically 1000 l s$^{-1}$). The catalysts used in the source interact with the gases, and a memory effect in the catalysts exists which is clearly observed in the experiments. Several hours of gas flow are needed after switching from natural hydrogen (mainly protium) to deuterium, or vice versa. The exchange of the two gases in the source is thus monitored and controlled. That protium should give the most prominent peak in the spectra under production conditions for D(0) is ruled out. If it did, other TOF peaks would also be identified as p(0), which is not the case. A small "contamination" by p in the D(0) cluster structure would further not be able to give the main peaks in our Table 3.

There exists one more possible origin to be discussed for the hypothetical protium in these experiments, and that is the fusion process D+D → p + T (branching, 50%). Such fusion processes have been shown to exist in this system [6]. However, the laser intensity in the experiments in Ref. [2] is not large enough to give intense laser-induced nuclear reactions, even if the limit to easily observed nuclear fusion is only a factor of three higher than the intensity used in [2]. When Ref. [2] was accepted, the results on fusion [6] were not published, which made it impossible for us to discuss this point. However, the laser intensity used in [2] is still judged to be too low to give a large density of protons and p(0) on the target, also based on other types of experiments. The arguments that the signal is clearly different using p(0) and D(0) and that a p(0) contamination would be easily observed are of course still valid. Thus, we conclude that an intense enough source of protons was not available in the experiments using D(0) in Ref. [2] which means that our analysis in Ref. [2] is valid.

## Acknowledgment
I want to thank Sveinn Olafsson for making me aware that Ref. [1] existed

# References

1. K. Hansen, Int. J. Mass Spectrom. 399-400 (2016) 51-52.
2. L. Holmlid, Int. J. Mass Spectrom. 352 (2013) 1- 8.
3. L. Holmlid, Int. J. Mass Spectrom. 351 (2013) 61-68.
4. P.U. Andersson and L. Holmlid, Int. J. Mass Spectrom. 310 (2012) 32-43.
5. F. Olofson and L. Holmlid, J. Appl. Phys. 111 (2012) 123502.
6. F. Olofson and L. Holmlid, Int. J. Mass Spectrom. 374 (2014) 33–38.

# Figure caption

Fig. 1. Data for $p^+$ KER (kinetic energy release) ejection from p(0) in Ref. [3], replotted as proposed in Ref. [1]. A relatively large KER is found, and an unrealistic small mass of 0.60 u instead of the known mass of 1 u. Thus the procedure used in Ref. [1] consistently gives a too low mass and is not valid.

Fig. 2. Tumbling rotational excitation for ultra-dense cluster $H_{2N}(0)$ by laser-induced Coulomb explosion. Each ellipse indicates a $H_2$ pair in the cluster. A longer cluster will be able to absorb a larger part of the KER as rotation. It was shown in Ref. [5] that the clusters stand vertically on a metal surface at large enough density.

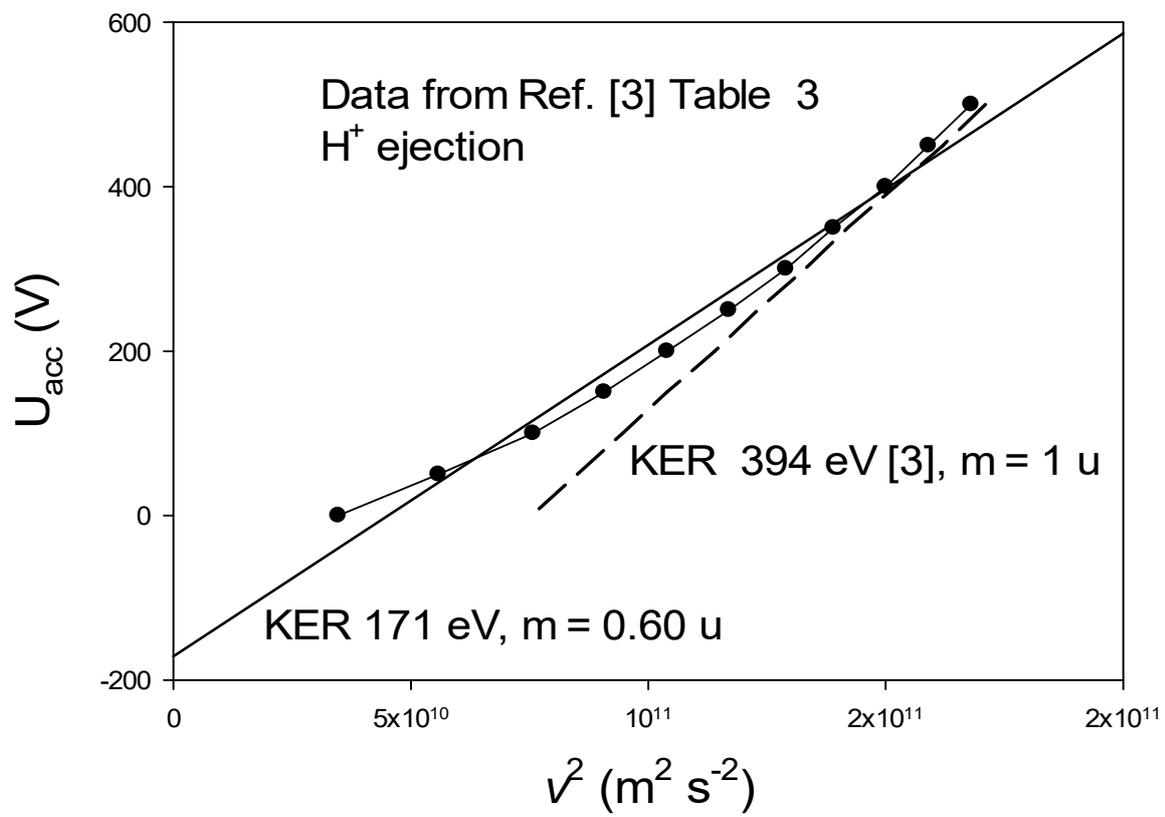

Fig. 1

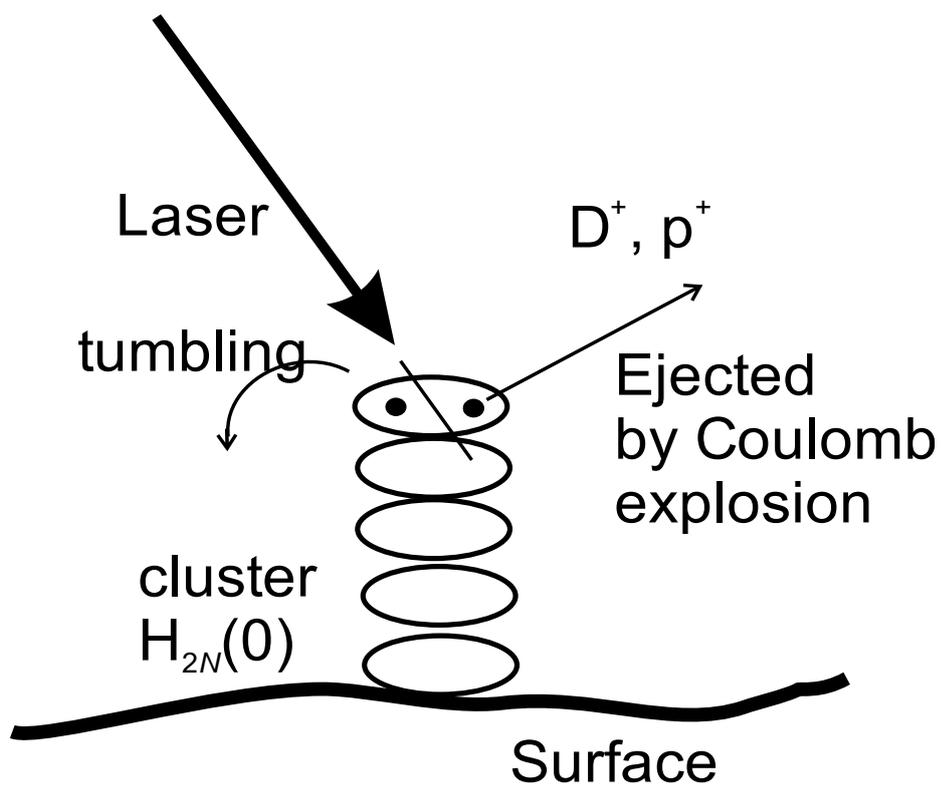

Fig. 2